\documentclass{sig-alternate}

\newcommand{\ignore}[1]{}
\usepackage{fancyhdr}
\usepackage[normalem]{ulem}
\usepackage[hyphens]{url}
\usepackage{balance}

\newcommand{\code}[1]{\texttt{#1}}

\usepackage{listings}
\usepackage{color}
\usepackage{algorithm, algorithmic}
\usepackage{blindtext}
\usepackage{cite}
\usepackage{color}   % \color
\usepackage{amssymb,amsmath}
\usepackage{comment} 
\usepackage{enumerate}
\usepackage{framed}
\usepackage{array}      % table tools
\usepackage{url}		% hyperlinks
\usepackage{balance}
\usepackage{xspace}
\usepackage{url}         % \url
\usepackage{multirow}
\usepackage{listings}
\usepackage{booktabs}
\usepackage{amsmath}
\usepackage{datetime}
\usepackage{subcaption}

\newcommand{\tool}{SSAM}

\fancypagestyle{firstpage}{
  \fancyhf{}
\setlength{\headheight}{50pt}

  \pagenumbering{arabic}
}  

\graphicspath{{figures/}{pics/}}
\DeclareGraphicsExtensions{.pdf,.jpeg,.png}

\title{Application-Driven Near-Data Processing for Similarity Search\vspace{-2ex}}

\author{Vincent T. Lee, Amrita Mazumdar, Carlo C. del Mundo, Armin Alaghi, Luis Ceze, Mark Oskin\\
        University of Washington\\
        {\{vlee2, amrita, cdel, armin, luisceze, oskin\}}@cs.washington.edu
}

\begin{document}
\pagenumbering{arabic}
\sloppy
\maketitle
\pagestyle{plain}

%%%%%% -- PAPER CONTENT STARTS-- %%%%%%%%

\begin{abstract}
\vspace*{0.20in}

Similarity search is a key to a variety of applications including content-based search for images and video, recommendation systems, data deduplication, natural language processing, computer vision, databases, computational biology, and computer graphics. 
At its core, similarity search manifests as k-nearest neighbors (kNN), a computationally simple primitive consisting of highly parallel distance calculations and a global top-k sort. 
However, kNN is poorly supported by today's architectures because of its high memory bandwidth requirements.

This paper proposes an application-driven near-data processing accelerator for similarity search: the Similarity Search Associative Memory (\tool{}).
By instantiating compute units close to memory, \tool{} benefits from the higher memory bandwidth and density exposed by emerging memory technologies.
We evaluate the \tool{} design down to layout on top of the Micron hybrid memory cube (HMC), and show that \tool{} can achieve up to two orders of magnitude area-normalized throughput and energy efficiency improvement over multicore CPUs; we also show \tool{} is faster and more energy efficient than competing GPUs and FPGAs.
Finally, we show that \tool{} is also useful for other data intensive tasks like kNN index construction, and can be generalized to semantically function as a high capacity content addressable memory.

\end{abstract}

\begin{comment}

\end{comment}

\section{Introduction}

Similarity search is a key computational primitive found in a wide range of applications, such as computational biology~\cite{knn-computational-biology}, computer graphics~\cite{knn-ray-tracing}, image and video retrieval~\cite{knn-video-search, knn-image-search}, image classification~\cite{nn-based-image-classification}, content deduplication~\cite{document_distances, dedup}, machine learning, databases~\cite{agrawal93}, data mining~\cite{top-10-data-mining}, and computer vision~\cite{tiny-images}. %, and data compression~\cite{mpeg-compression}. 
While much attention has been directed towards accelerating feature extraction techniques like convolutional neural networks~\cite{diannao}, there has been relatively little work focused on accelerating the task that follows: taking the resulting feature vectors and searching the vast corpus of data for similar content. 
In recent years, the importance and ubiquity of similarity search has increased dramatically with the explosive growth of visual content: users shared over 260 billion images on Facebook in 2010~\cite{facebook-photos}, and uploaded over 300 hours of video on YouTube every minute in 2014~\cite{youtube-statistics}. 
This volume of visual data is only expected to continue growing exponentially~\cite{rebooting-the-it-revolution}, and has motivated search-based graphics and vision techniques such as visual memex~\cite{visual-memex-nips09}, 3D reconstruction~\cite{3D-reconstruction}, and cross-domain image matching~\cite{data-driven-similarity}. %image completion/editing~\cite{dreambit}. 

Similarity search manifests as a simple algorithm: k-nearest neighbors (kNN).
At a high level, kNN is an approximate associative computation which tries to find the most similar content with respect to the query content.
At its core, kNN consists of many parallelizable distance calculations and a single global top-k sort, and is often supplemented with indexing techniques to reduce the volume of data that must be processed.
While computationally very simple, kNN is notoriously memory intensive on modern CPUs and heterogeneous computing substrates making it challenging to scale to large datasets.
In kNN, distance calculations are cheap and abundantly parallelizable across the dataset, but moving data from memory to the computing device is a significant bottleneck. 
Moreover, this data is used only once per kNN query and discarded since the result of a kNN query is only a small set of identifiers. 
Batching requests to amortize this data movement has limited benefits as time-sensitive applications have stringent latency budgets.
Indexing techniques such as kd-trees~\cite{kd-tree}, hierarchical k-means clustering~\cite{flann}, and locality sensitive hashing~\cite{lsh} are often employed to reduce the search space but trade reduced search accuracy for enhanced throughput.
Indexing techniques also suffer from the \textit{curse of dimensionality}~\cite{Indyk:1998:ANN:276698.276876}; in the context of kNN, this means indexing structures effectively degrade to linear search for increasing accuracy targets.

%Also, the \textit{curse of dimensionality}~\cite{Indyk:1998:ANN:276698.276876} indicates that the hierarchical index data structures used in kNN, such as kd-trees and k-means clustering, require significantly more linear search as dimensionality grows~\cite{Datar:2004:LHS:997817.997857, gionis1999similarity, weber1998quantitative}, further increasing pressure on memory bandwidth and distance computation engines. 

Because of its significance, generality, parallelism, underlying simplicity, and small result set, kNN is an ideal candidate for near-data processing.
The key insight is that a small accelerator can reduce the traditional bottlenecks of kNN by applying orders of magnitude data reduction near memory, substantially reducing the need for data movement.
While there have been many attempts at processing-in-memory (PIM) in the past~\cite{McKee95,iram,flexram,DIVA,impulse,activepages}, much of prior work suffered from DRAM technology limitations.
Logic created in DRAM processes was too slow, while DRAM implemented in logic processes suffered from poor retention and high power demands; attempts at hybrid processes~\cite{sa-27} result in the worst of both.
PIM architectures are more appealing today with the advent of die-stacked memory technology which enables the co-existence of an efficient DRAM layer \emph{and} efficient logic layer~\cite{HMC}.

We propose Similarity Search Associative Memory (\tool{}) which integrates a programmable accelerator into a die-stacked memory module.
Semantically, a \tool{} takes a query as input and returns the top-k closest neighbors stored in memory as output.
We evaluate the performance and energy efficiency gains of \tool{} by implementing, synthesizing, and simulating the design down to layout.
We then compare \tool{} against current CPUs, GPUs, and FPGAs, and show that it can achieve better area-normalized throughput and energy efficiency.

Our paper makes the following contributions:
{
\begin{itemize}
\setlength{\itemsep}{0.05cm}
\item A characterization of state-of-the-art k-nearest neighbors including both application-level and architectural opportunities that justify acceleration.
\item An application-driven codesign of a near memory vector processor-based accelerator architecture with hardware support for similarity search on top of Hybrid Memory Cube (HMC).
\item Instruction extensions to leverage hardware units to accelerate similarity search.
\end{itemize}
}

The rest of the paper is organized as follows.
Section~\ref{sec:characterization} introduces and characterizes the kNN algorithm. 
Section~\ref{sec:architecture} describes the \tool{} architecture and the hardware/software interface. 
Section~\ref{sec:methodology} outlines evaluation methodology, and Section~\ref{sec:evaluation} presents evaluation results.
Section~\ref{sec:discussion} discusses the impact of these results on different application areas and design points.
Finally, Section~\ref{sec:related-works} discusses related work.  

%%%%%%%%%%%%
%%%%
%%%% Bit bucket.  Do not put active content in here.
%%%%
%%%%%%%%%%%%

\begin{comment}

\end{comment}

\section{Characterization of kNN} \label{sec:characterization}
\label{sec:background}
We now introduce and characterize the kNN algorithm pipeline and indexing techniques, and highlight the application-level and architectural opportunities for acceleration.

\subsection{Case study: content-based search}

A typical kNN software application pipeline for content-based search (Figure~\ref{visual-search}) has five stages: feature extraction, feature indexing, query generation, k-nearest neighbors search, and reverse lookup.
In \textit{feature extraction} (Figure~\ref{visual-search}a), the raw multimedia corpus is converted into an intermediary feature vector representation.
Feature vectors may represent pixel trajectories in a video, word embeddings of a document, or shapes in an image~\cite{dense-trajectories, document_distances, sift}, and are extracted using feature descriptors or convolutional neural networks~\cite{sift, surf, gist, alexnet, googlenet, resnet}.
While feature extraction is an important component of this pipeline, it only needs to be performed once for the dataset and can be done offline; a significant portion of work has also shown feature extraction can be achieved efficiently~\cite{hauswald15asplos, du15, hauswald15isca, chen14, diannao}.
In \textit{indexing} (Figure~\ref{visual-search}b), feature vectors from feature extraction are organized into data structures (discussed in Section~\ref{sec:indexing}).
At query time, these data structures are used to quickly prune the search space; intuitively, these data structures should be able to reduce the search time from linear to logarithmic in the size of the data.
Indexing, like feature extraction, can be performed offline and away from the critical path of the query.

While feature extraction and indexing can be performed offline, the \textit{query generation} stage (Figure~\ref{visual-search}c) of the search pipeline occurs online.
In query generation, a user uploads a multimedia file (image, video, etc.) and requests similar content back.
The query runs through the same feature extractor used to create the database before being passed to the search phase.
Once a query is generated, the \textit{k-nearest neighbors} stage (Figure~\ref{visual-search}d) attempts to search for the most similar content in the database.
The kNN algorithm consists of many highly parallelizable distance calculations and a global top-k sort; indexing structures may also be employed to prune the search space but trade accuracy for performance.
The similarity metric employed by the distance calculation often depends on the application, but common distance metrics include Euclidean distance, Hamming distance~\cite{gionis99, LSH_Lazebnik, LDAHash, Torralba08smallcodes, wang12, spectral-hashing, k-means-hashing, hamming-metric-learning}, cosine similarity~\cite{tagspace}, and learned distance metrics~\cite{distance-metric-learning}.
The final step in the pipeline is \textit{reverse lookup} where the resulting $k$ nearest neighbors are mapped to their original database content.
The resulting media is then returned to the user.

{
\begin{figure*}[!ht]
\centering
\includegraphics[width=\linewidth]{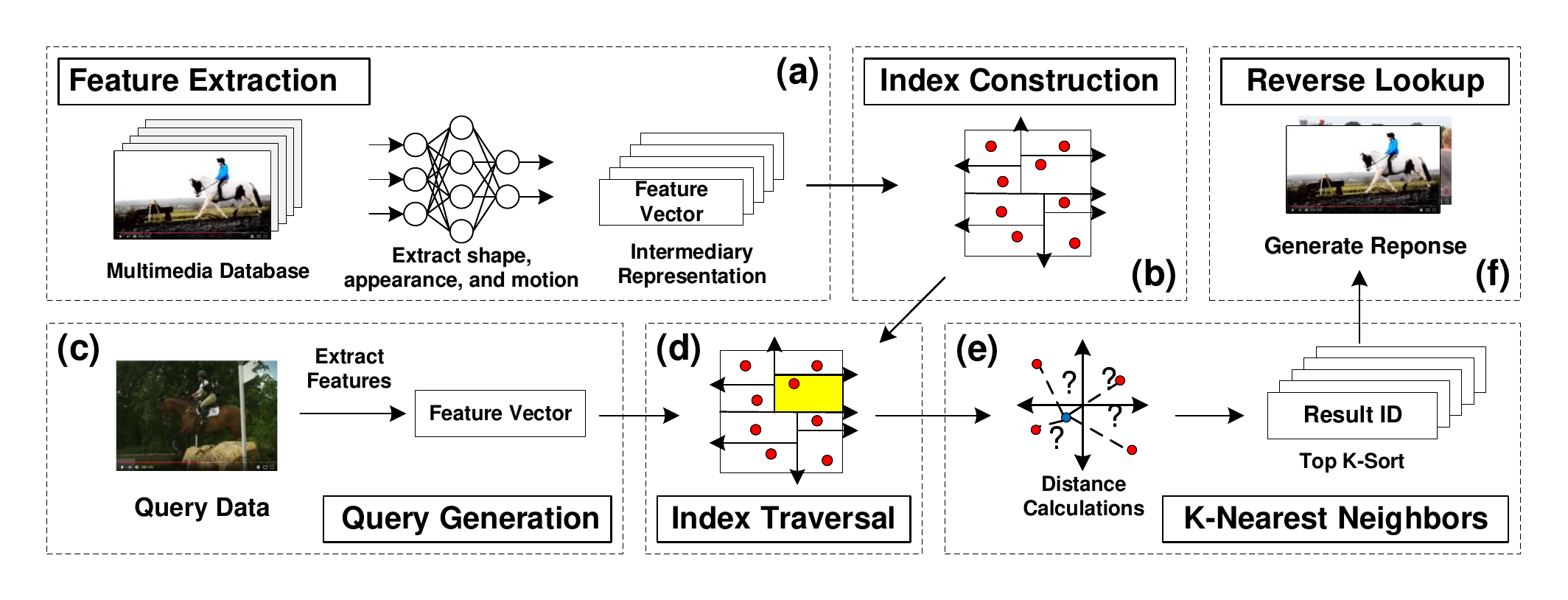}
\caption{Software application pipeline for similarity search. Feature extraction and indexing is done offline. (a) feature extraction, (b) feature indexing, (c), query generation, (d) index traversal (e) k-nearest neighbor search, (f) reverse lookup.}
\label{visual-search}
\end{figure*}
}

\subsection{Typical workload parameters}

Prior work shows the feature dimensionality for descriptors such as Speeded Up Robust Feature (SURF)~\cite{surf}, word embeddings~\cite{glove-dataset}, Scale Invariant Feature Transform (SIFT)~\cite{sift}, GIST descriptors~\cite{gist}, AlexNet~\cite{jia2014caffe}, and ResNet~\cite{resnet} ranges from 64 to 4096 dimensions.
For higher dimensional feature vectors, it is common to apply techniques such as principal component analysis to reduce feature dimensionality to tractable lengths~\cite{pca}.
The number of nearest neighbors $k$ for an array of search applications has been shown to range from 1 (nearest neighbor) up to 20~\cite{sift, document_distances, motion_planning, how-many-results, youtube-statistics}.
Each kNN algorithm variant also has a number of additional parameters such as indexing technique, distance function, bucket size, index-specific hyperparameters, and hardware specific optimizations.

To simplify the characterization, we limit our initial evaluation to Euclidean distance and three real world datasets: the Global Vectors for Word Representations (GloVe) dataset~\cite{glove-dataset}, the GIST dataset~\cite{gist-dataset}, and the AlexNet dataset~\cite{flickr}.
The GloVe dataset consists of 1.2 million word embeddings extracted from Twitter tweets and the GIST dataset consists of 1 million GIST feature vectors extracted from images.
We also constructed an AlexNet dataset by taking 1 million images from the Flickr dataset~\cite{flickr} and applying AlexNet~\cite{alexnet} to extract the feature vectors.
For each dataset, we separate it into a ``training'' set used to build the search index, and a ``test'' set used as the queries when measuring application accuracy.
Exact dataset parameters used for our characterization and evaluation are shown in Table~\ref{knn-workloads}.

{
\begin{table}[!ht]
\centering
\caption{Evaluated kNN workload parameters}
\begin{tabular}{@{}ccccc@{}} \toprule
\textbf{Workload} & \begin{tabular}[x]{@{}c@{}}\textbf{Dataset}\\\textbf{Vectors}\end{tabular} & \textbf{Queries} & \begin{tabular}{@{}c@{}}\textbf{Dimen-}\\\textbf{sions}\end{tabular} & 
\begin{tabular}{@{}c@{}}\textbf{Neigh-}\\\textbf{bors}\end{tabular} \\ \midrule
GloVe & 1183514 & 10000 & 100 & 6 \\
GIST & 1000000 & 1000 & 960 & 10 \\
AlexNet & 1000000 & 1000 & 4096 & 16 \\ \bottomrule
\end{tabular}
\label{knn-workloads}
\end{table}
}

\subsection{Approximate kNN algorithms tradeoffs}

\label{sec:indexing}

We now characterize three canonical indexing techniques employed by approximate kNN algorithms: kd-trees, hierarchical k-means, and multi-probe locality sensitive hashing (MPLSH).
Indexing techniques employ hierarchical data structures which are traversed at query time to prune the search space.
In kd-trees, the index is constructed by randomly cutting the dataset by the top-N vector dimensions with highest variance~\cite{randomized-kd-tree}.
The resulting index is a tree data structure where each leaf in the tree contains a \textit{bucket} of similar vectors; the depth of the bucket depends on how tall the tree is limited to be.
Queries which traverse the index and end up in the same bucket should be similar; multiple parallel trees are often used in parallel with different cut orders.
Multiple leaves in the tree can be visited to improve the quality of the search; to do this, the traversal employs \textit{backtracking} to check additional ``close by'' buckets in a depth first search-like fashion.
A user-specified bound typically limits the number of additional buckets visited when backtracking.

Similarly, in hierarchical k-means the dataset is partitioned recursively based on k-means cluster assignments to form a tree data structure~\cite{flann}.
Like kd-tree indices, the height of the tree is restricted, and each leaf in the tree holds a bucket of similar vectors which are searched when a query reaches that bucket; backtracking is also used to expand the search space and search ``close by'' buckets.

Finally, MPLSH constructs a set of hash tables where each hash location is associated with a bucket of similar vectors~\cite{MPLSH}.
In MPLSH, hash functions are designed to intentionally cause hash collisions to map similar vectors to the same bucket.
To improve accuracy, MPLSH applies small perturbations to the hash result to create additional probes into the same hash table to search ``close by'' hash partitions.
In our evaluation, we use hyperplane MPLSH (HP-MPLSH) which cuts the space into random hyperplanes and set the number of hash bits or hyperplane cuts to 20.

Each of these approximate kNN algorithms trade accuracy for enhanced throughput.
In kNN, accuracy is defined as $S_E \cap S_A / |S_E|$ where $S_E$ is the true set of neighbors returned by exact floating point based linear kNN search, and $S_A$ is the set of neighbors returned by approximate kNN.
In general, searching more of the dataset improves search accuracy for indexing techniques.
To quantify the accuracy of indexing structures, we benchmark the accuracy and throughput of indexing techniques for the GloVe, GIST, and AlexNet datasets.
We use the Fast Library for Approximate Nearest Neighbors (FLANN)~\cite{flann} to benchmark kd-trees and hierarchical k-means, and Fast Lookups for Cosine and Other Nearest Neighbors Library (FALCONN)~\cite{falconn} to benchmark HP-MPLSH.
For kd-trees and hierarchical k-means we vary the number of leaf nodes or buckets in the tree that backtracking will check, while for HP-MPLSH we increase the number of probes used per hash table.
Each of these modifications effectively increases the fraction of the dataset searched per query and lowers overall throughput.

The resulting throughput versus accuracy curves are shown in Figure~\ref{char:accuracy} for single threaded implementations.
In general, our results show indexing techniques can provide up to 170$\times$ throughput improvement over linear search while still maintaining at least 50\% search accuracy, but only up to 13$\times$ in order to achieve 90\% accuracy.
Past 95-99\% accuracy, we find that \textit{indexing techniques effectively degrade to linear search} (blue solid line).
More importantly, our results show there is a significant opportunity for also accelerating approximate kNN techniques.
Hardware acceleration of approximate kNN search can \textit{either increase throughput at iso-accuracy} by simply speeding up the computation or \textit{increase search accuracy at iso-latency} by searching larger volumes of data.

{
\setlength{\floatsep}{-10pt}
\begin{figure*}[!ht]
\centering
\includegraphics[width=\linewidth]{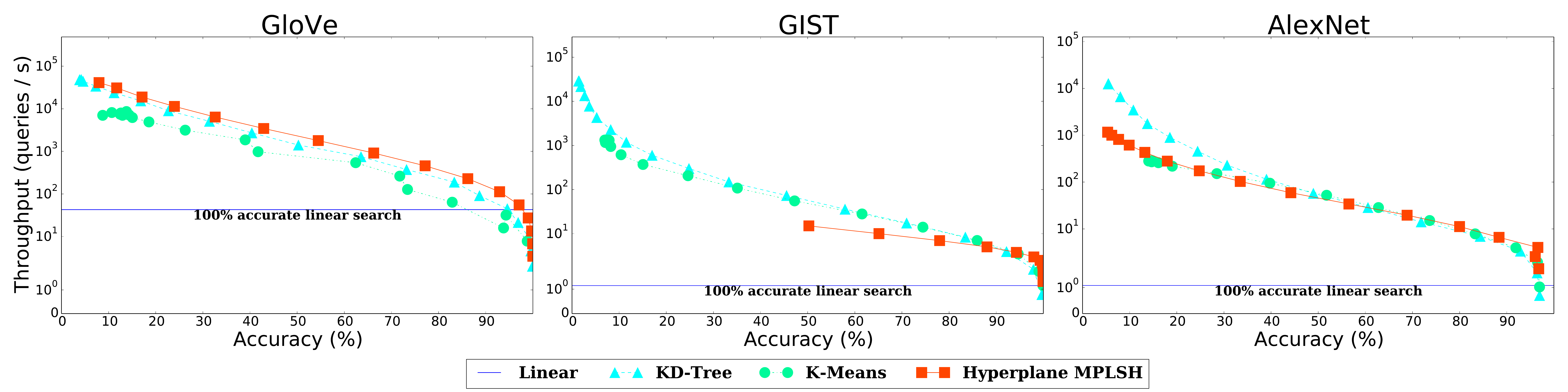}
\caption{Approximate kNN algorithms tradeoff accuracy for throughput (up and to the right is better).}
\label{char:accuracy}
\end{figure*}
}

\subsection{Alternative numerical representations and distance metrics}

We now briefly discuss the space of numerical representations and distance metrics used in kNN search.

\noindent\textbf{Fixed-Point Representations}: Fixed-point arithmetic is much cheaper to implement in hardware than floating point units.
To evaluate whether floating point is necessary for kNN, we converted each dataset to a 32-bit fixed-point representation and repeated the throughput versus accuracy experiments.
Overall, we find \textit{there is negligible accuracy loss} between 32-bit floating-point and 32-bit fixed-point data representations.

\noindent\textbf{Hamming Space Representations}: A large body of recent work has shown that Hamming codes can be an effective alternative for Euclidean space representations~\cite{k-means-hashing, itq, hamming-metric-learning, spectral-hashing, semi-supervised-hashing, small-codes-databases, gionis99}.
Binarization techniques trade accuracy for higher throughput since precision is lost by binarizing floating point values but throughput increases since the dataset size is smaller; binarization also enables Hamming distance calculations which are cheaper to implement in hardware.
In practice, carefully constructed Hamming codes have been shown to achieve excellent results~\cite{lin16}.

\noindent\textbf{Alternative Distance Metrics}: While the canonical distance metric for kNN is the Euclidean norm, there still exist a wide variety of alternative distance metrics.
Such alternative metrics include Manhattan distance, cosine similarity, Chi squared distance, Jaccard similarity, and learned distance metrics~\cite{distance-metric-learning}.

\subsection{Architectural Characterization}

To more concretely quantify the architectural behaviors of kNN variants, we instrumented the baselines presented earlier using the Pin~\cite{pin-tool} instruction mix tool on an Intel i7-4790K CPU. 
Table~\ref{knn-profile} shows the instruction profile for linear, kd-tree, k-means, and MPLSH based algorithms respectively.
Recall that linear search performance is still valuable since higher accuracy targets reduce to linear search; in addition, approximate algorithms still use linear search to scan buckets of vectors at the end of their traversals.
As expected, the instruction profile shows that vector operations and extensions are important for kNN workloads due to the many vector-parallel distance calculations.
In addition, the high percentage of memory reads confirms that the computation has high data movement demands.
Approximate kNN techniques like KD-trees and MPLSH exhibit less skew towards vectorized instructions but still exhibit similar memory intensive behavior and show vectorization is valuable.

\begin{table}
\centering
\caption{Architectural behavior profiles of kNN algorithms for GloVe dataset.}
\label{knn-profile}
\begin{tabular}{@{}cccc@{}} 
\toprule
Algorithm      & 
\begin{tabular}{@{}c@{}}AVX/SSE\\Inst. (\%) \end{tabular} & 
\begin{tabular}{@{}c@{}}Mem. \\Reads (\%)\end{tabular} & 
\begin{tabular}{@{}c@{}}Mem. \\Writes (\%)\end{tabular} \\ \midrule
Linear         & 54.75           & 45.23         & 0.44           \\
KD-Tree        & 28.75           & 31.60         & 10.21          \\
K-Means        & 51.63           & 44.96         & 1.12           \\
MPLSH          & 18.69           & 31.53         & 14.16          \\
%KD-Tree (50\%) & 28.15           & 32.99         & 13.87          \\
%KD-Tree (90\%) & 28.75           & 31.60         & 10.21          \\
%K-Means (50\%) & 51.63           & 44.63         & 1.57           \\
%K-Means (90\%) & 51.63           & 44.96         & 1.12           \\
%MPLSH (50\%)   & 18.69           & 31.53         & 14.16          \\
%MPLSH (90\%)   & 17.12           & 30.84         & 15.49          \\
\bottomrule
\end{tabular}
\end{table}

%%%%%%%%%%%%%%%%%%%%%%%%%%%%%%%%%%%%%%%%%%%%%%%%%%%%%%%
% Bit bucket
%%%%%%%%%%%%%%%%%%%%%%%%%%%%%%%%%%%%%%%%%%%%%%%%%%%%%%%

\begin{comment}

\begin{table}[!ht]
\centering
\caption{kNN parameter space}
\label{knn-parameter-space}
\begin{tabular}{@{}cc@{}} \toprule
\textbf{Parameter} & \textbf{Typical Values} \\ \midrule
Indexing Structure & \begin{tabular}{@{}c@{}}None, KD-Tree, K-Means,\\MPLSH, Other\end{tabular} \\ \hline
Neighbors $k$ & 1-20 \\ \hline
Dimensions $d$ & 64-4096 \\ \hline
Bucket Size $B$ & Integer \\ \hline
Distance Metric & \begin{tabular}{@{}c@{}}Euclidean, Hamming, cosine,\\Chi squared, learned metrics\end{tabular}\\ \hline
\begin{tabular}{@{}c@{}}Numerical\\Representation\end{tabular} & \begin{tabular}{@{}c@{}}32-bit float, int, binary,\\reduced precision fixed point\end{tabular} \\ \bottomrule
\end{tabular}
\end{table}

\end{comment}

\section{\tool{} Architecture} \label{sec:architecture}

Based on the characterization results in Section~\ref{sec:background}, it is clear that similarity search algorithms (1) are an ideal match for vectorized processing units, and (2) can benefit from higher memory bandwidth to better support its data intensive execution phases.
We now present our application-guided \tool{} module and accelerator architecture which exploits near-data processing and specialized vector compute units to address these bottlenecks.

\subsection{System integration and software interface}

\tool{} is a memory module that integrates into a typical system as a memory module similar to existing DRAM as shown in Figure~\ref{system-architecture}.
A host processor interfaces with an \tool{} module similar to how it interacts with a DRAM memory module.
The host processor is connected to each \tool{} module over a communication bus; additional communication links are used if multiple \tool{}-enabled modules are instantiated.
Since HMC modules can be composed together, these additional links and \tool{} modules allows us to scale up the capacity of the system.
A typical system may also have multiple host processors (not shown) as the number of \tool{} modules that the system must maintain increases.

{
\begin{figure}[!ht]
\centering
\includegraphics[width=\linewidth]{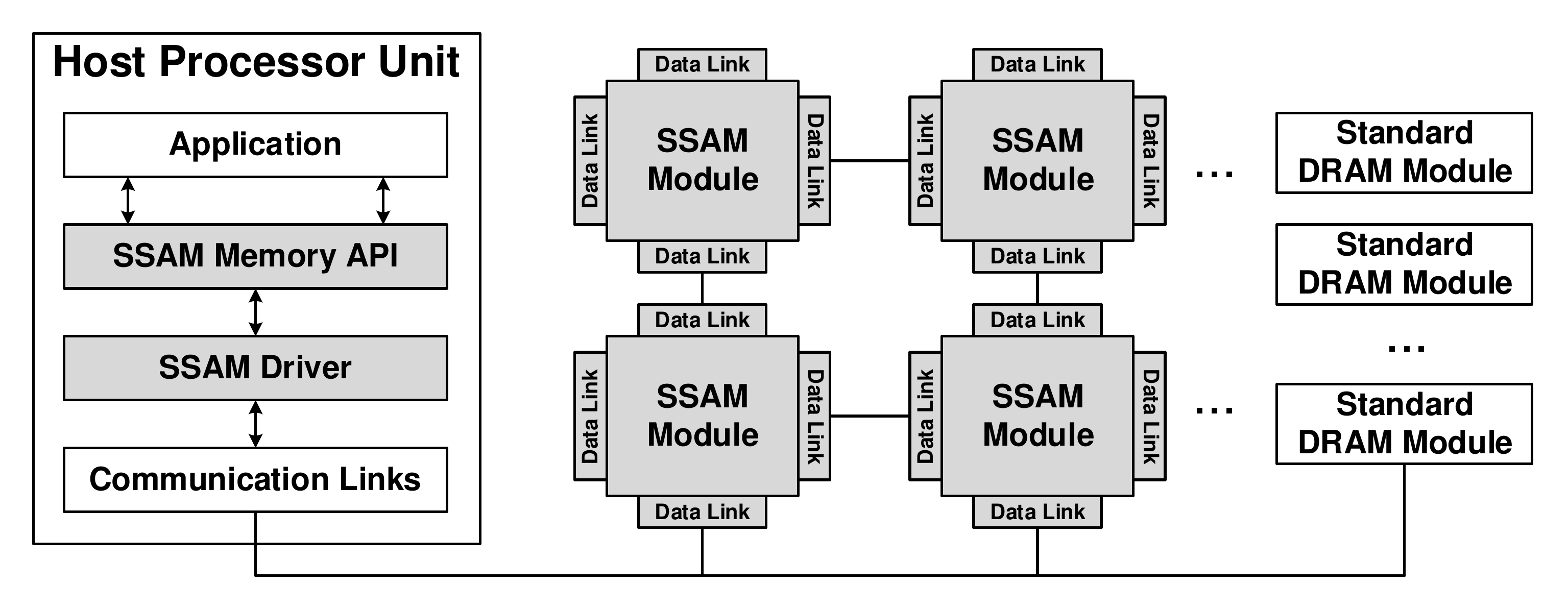}
\caption{\tool{} system integration (modifications in grey). \tool{} modules replace or coexist with standard DRAM modules.}
\label{system-architecture}
\end{figure}
}

To abstract the lower level details of \tool{}s away from the programmer, we assume a driver stack exposes a minimal memory allocation API which manages user interaction with \tool{}-enabled memory regions.
An \tool{}-enabled memory region is defined as a special part of the memory space which is physically backed by an \tool{} instead of a standard DRAM module.
A sample programming interface of how one would use \tool{}-enabled memory regions is shown in Figure~\ref{example_program}.
\tool{}-enabled memory regions would be tracked and stored in a free list similar to how standard memory allocation is implemented in modern systems.
Allocated \tool{} memory regions come with a set of special operations that allow the user to set the indexing mode, in additional to handling standard memory manipulation operations like \code{memcpy}.
Similar to the CUDA programming model, we use analogous memory and execution operations to operate \tool{}-enabled memory.
Pages with data subject to \tool{} queries are pinned (not subject to swapping by the OS). %, and \tool{} queries are assumed to not cross page boundaries.

%\bluecomment{TODO: cite all the other near data processing work that resolves other issues like virtualization, coherence, etc.}

{ \small
\begin{figure}[!ht]
\lstset{language=C,
                basicstyle=\ttfamily,
                keywordstyle=\color{blue}\ttfamily,
                stringstyle=\color{red}\ttfamily,
                commentstyle=\color{black}\ttfamily,
                morecomment=[l][\color{magenta}]{\#}
}

\begin{lstlisting}[language=C]
// Example program using SSAM
int * knn(int * query, int *dataset,
   size_t length, size_t dims, int k) {
  //allocate buffer of SSAM memory
  int * nbuf = nmalloc(length * dims);
  nmode(nbuf, LINEAR);
  nmemcpy(nbuf, dataset, length * dims
       * sizeof(int));
  nbuild_index(nbuf, params = NULL);  
  nwrite_query(nbuf, query);
  //execute kNN search
  nexec(nbuf);
  int * result = nread_result(nbuf);
  nfree(nbuf);
  return result;
}
\end{lstlisting}
\caption{Example program using \tool{}-enabled memory regions. Lower level hardware configuration details are abstracted away from the programmer.}
\label{example_program}
\end{figure}
}

\subsection{\tool{} architecture and hybrid memory cube}

The \tool{} architecture is built on top of a Hybrid Memory Cube 2.0 (HMC) memory substrate~\cite{hmc-2.0} to capitalize on enhanced memory bandwidth.
The HMC is a die-stacked memory architecture composed of multiple DRAM layers and a compute layer.
The DRAM layers are vertically partitioned into a number of \textit{vaults} (Figure~\ref{ncam-accelerator}a).
Vaults are each accessed via a vault controller which reside on a top-level compute layer.
In HMC 2.0, the module is partitioned into a maximum of 32 vaults (only 16 are shown), where each vault controller operates at 10 GB/s yielding an aggregate internal memory bandwidth of 320 GB/s.
The HMC architecture also is composed of four external data links (240 GB/s external bandwidth) which send and receive information to the host processor or other HMC modules.
These external data links allow one or more HMC modules to be composed to effectively form a larger network of \tool{}s if data exceeds the capacity of a single \tool{} module.

Our \tool{} architecture leverages the existing HMC substrate and introduces a number of \tool{} accelerators to handle the kNN search.
These \tool{} accelerators are instantiated on the compute layer next to existing vault controllers as shown in Figure~\ref{ncam-accelerator}b.
\tool{} accelerators are further decomposed into processing units (Figure~\ref{ncam-accelerator}d).
To fully harness the bandwidth available, we replicate processing units to fully use the memory bandwidth by measuring the peak bandwidth needs of each processing unit across all indexing techniques.
For kNN, we expect to achieve near optimal memory bandwidth since almost all data accesses to memory are large contiguous blocks such as bucket scans and data structures, which are contiguously allocated in memory.
Our modifications are made orthogonal to the functionality of the HMC control logic so that the HMC module can still operate as a standard memory module (i.e. acceleration logic can be bypassed).
Our processing units do not implement a full cache hierarchy since there is little data reuse outside of the query vector and indexing data structure per query.
Unlike GPUs cores, processing units are not restricted to operating in lockstep and multiple different indexing kernels can coexist on each \tool{} module.
Finally, we do not expect external data links to become a bottleneck as a vast majority of the data movement occurs within \tool{} modules themselves.
As a result, we only expect the communication network between the host processors and \tool{} units to consist of kNN results which are a fraction of the original dataset size, and configuration data.

{
\begin{figure*}[!ht]
\centering
\includegraphics[width=\linewidth]{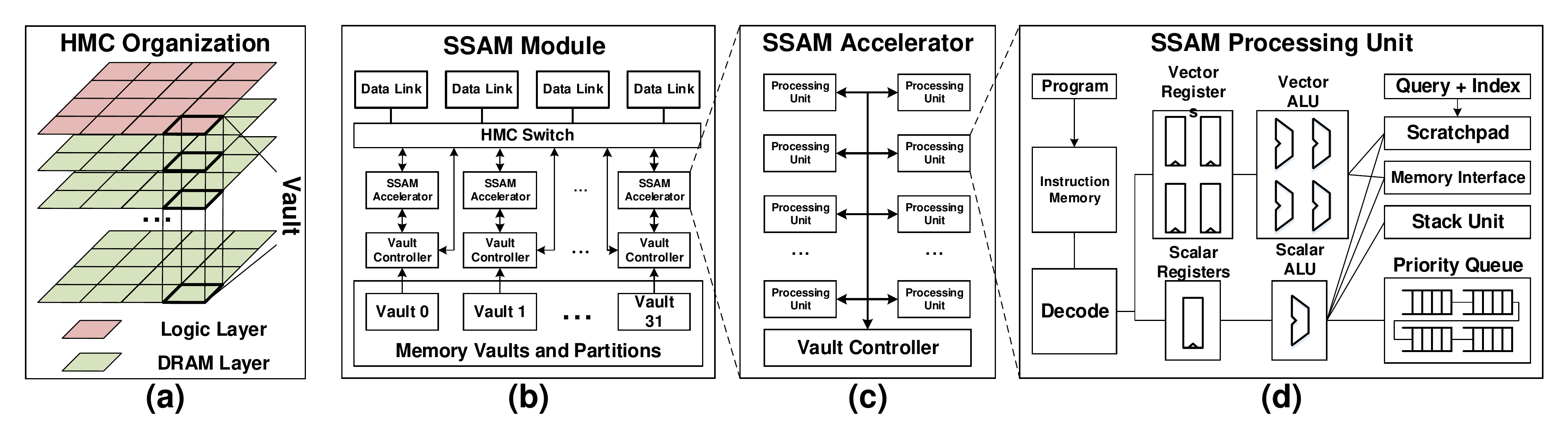}
\caption{\tool{} accelerator architecture. (a) HMC die organization (only 16 vaults shown, HMC 2.0 has 32), (b) \tool{} logic layer organization, (c) \tool{} accelerator organization, (d) processing unit microarchitecture.}
\label{ncam-accelerator}
\end{figure*}
}

\subsection{Processing unit architecture}

Each processing unit consists of a fully integrated scalar and vector processing unit similar to~\cite{russell78} but are augmented with several instructions and hardware units to better support kNN.
Fully-integrated vector processing units are naturally well-suited for accelerating kNN distance calculations because they are (1) able to exploit the abundant data parallelism in kNN and (2) well-suited for streaming computations.
Using vector processing units also introduces flexibility in the types of distance calculations that can be executed.
The scalar unit is better suited for executing index traversals which are sequential in nature, and provides flexibility in the types of indexing techniques that can be employed.
Vector units on the other hand are better suited for high throughput data parallel distance calculations in kNN.
We use a single instruction stream to drive both the scalar and vector processing units since at any given time a processing unit will only be performing either distance calculations or index traversals in kNN.
For our evaluation, we perform a design sweep over several different vector lengths: 2, 4, 8, and 16.
We find that 32 scalar registers, and 8 vector registers are sufficient to support our kNN workloads.
Finally, we use forwarding paths between pipeline stages to implement chaining of vector operations.

We also integrate several hardware units that are useful for accelerating similarity search.
First, we introduce a \textit{priority queue unit} implemented using a shift register architecture proposed in~\cite{shift-priority-queue}, and is used to perform the sort and global top-k calculations.
For our \tool{} design, priority queues are 16 entries deep.
We opt to provide a hardware priority queue instead of a software one since the overhead of a priority queue insert becomes non-trivial for shorter vectors.
Because of its modular design, the priority queues can be chained to support higher $k$ values; likewise, priority queues in the chain can also be disabled if they are not needed.
Second, we introduce a small hardware stack unit instantiated on the scalar datapath to aid kNN index traversals.
The stack unit is a natural choice to facilitate backtracking when traversing hierarchical index structures.
Finally, we integrate a 32 KB scratchpad to hold frequently accessed data structures, such as the query vector and indexing structures.
We find that a modestly sized scratchpad memory is sufficient for kNN since the only heavily reused data are the query vectors and indices (data vectors are scanned and immediately discarded).

Unlike conventional scalar-vector architectures, we introduce several new instructions to exercise new hardware units for similarity search.
First, we introduce priority queue insert (\code{PQUEUE\_INSERT}), load (\code{PQUEUE\_LOAD}), and reset (\code{PQUEUE\_RESET}) instructions which are used to manipulate the hardware priority queue.
The \code{PQUEUE\_INSERT} instruction takes two registers and inserts them into the hardware priority queue as an \code{(id, value)} tuple.
The \code{PQUEUE\_LOAD} instruction reads either the \code{id} or \code{value} of a tuple in the priority queue at a designated queue position, while the \code{PQUEUE\_RESET} clears the priority queue.
We also introduce a scalar and vector 32-bit \textit{fused xor-population count} instruction (\code{SFXP} and \code{VFXP}) which is similar to a fused multiply add instruction.
The \code{FXP} instruction is useful for cheaply implementing Hamming distance calculations and assumes that each 32-bit word is 32 dimensions of a binary vector.
The \code{FXP} instruction is also cheap to implement in hardware since the XOR only adds one additional layer of logic to the population count hardware.
Finally, we introduce a data prefetch instruction \code{MEM\_FETCH} since the linear scans through buckets of vectors exhibit predictable contiguous memory access patterns.

{
\begin{table*}
\centering
\caption{Processing unit instruction set. (S/V) are scalar and vector instructions. (S) instructions are scalar only.}
\label{ncam_isa}
\begin{tabular}{@{}ll@{}} \toprule
\textbf{Type} & \textbf{Instruction} \\ \midrule
Arithmetic (S/V) & \code{ADD}, \code{SUB}, \code{MULT}, \code{POPCOUNT}, \code{ADDI}, \code{SUBI}, \code{MULTI} \\
Bitwise/Shift (S/V) & \code{OR}, \code{AND}, \code{NOT}, \code{XOR}, \code{ANDI}, \code{ORI}, \code{XORI}, \code{SR}, \code{SL}, \code{SRA} \\ 
Control (S) & \code{BNE}, \code{BGT}, \code{BLT}, \code{BE}, \code{J} \\
Stack Unit (S) & \code{POP}, \code{PUSH} \\
Register Move/Memory Instructions (S/V) & \code{SVMOVE}, \code{VSMOVE}, \code{MEM\_FETCH}, \code{LOAD}, \code{LOAD}, \code{STORE} \\ \midrule
\textbf{New \tool{} Instructions} & \textbf{\code{(S)PQUEUE\_INSERT}}, \textbf{\code{(S)PQUEUE\_LOAD}}, \textbf{\code{(S)PQUEUE\_RESET}}, \textbf{\code{(S/V)FXP}}  \\ \bottomrule
\end{tabular}
\end{table*}
}

\begin{comment}
\begin{table}
\centering
\caption{Processing unit design parameters}
\begin{tabular}{@{}ll@{}} \toprule
\textbf{Design Parameter} & \textbf{Value} \\ \midrule
Vector Length & 2-16 \\
Scalar Registers & 32 \\
Vector Registers. & 8 \\
Scratchpad Size & 32 KB \\
Maximum Priority Queue Depth & 16 \\
Instruction Memory Size & 2 KB \\
Stack Unit Depth & 20 \\ \bottomrule
\end{tabular}
\label{pu-parameters}
\end{table}
\end{comment}

\subsection{\tool{} configuration}

We assume that the host processor driver stack is able to communicate with each \tool{} to initialize and bring up \tool{} devices using a special address region dedicated to operating \tool{}s.
Execution binaries are written to instruction memories on each processing unit and can be recompiled to support different distance metrics, indexing techniques, and kNN parameters.
In addition, any indexing data structures are also written to the scratchpad memory or larger DRAM prior to executing any queries on \tool{}s.
Any large data structures such as hash function weights in MPLSH or centroids in k-means are stored in \tool{} memory since they are larger and experience limited reuse.
If hierarchical indexing structures such as kd-trees or hierarchical k-means do not fit in the scratchpad, they are partitioned such that the top half of the hierarchy resides in scratchpad, and the bottom halves are dynamically loaded to the scratchpad from DRAM as needed during execution.
A small portion of the scratchpad is also allocated for holding the query vector; this region is continuously rewritten as an \tool{} services queries.
If an kNN query must touch multiple vaults, the host processor broadcasts the search across \tool{} processing units and performs the final set of global top-k reductions on the host processor.
Finally, if \tool{} capabilities are not needed, the host processor can disable the \tool{} accelerator logic so that it operates simply as a standard memory.

%%%%%%%%%%%%%%%%%%%%%%%%%%%%%%%%%%%%%%%%%%%%%%%%%%%
% Bit Bucket
%%%%%%%%%%%%%%%%%%%%%%%%%%%%%%%%%%%%%%%%%%%%%%%%%%%

\begin{comment}

\end{comment}

\section{Evaluation Methodology} \label{sec:methodology}

We now outline our evaluation methodology used to compare and contrast \tool{}s with competing CPUs, GPUs, and FPGAs shown in Table~\ref{evaluated-platforms}.
To provide fair energy efficiency and performance measurements, we normalize each platform to a 28 nm technology process.

{
\begin{table}
\centering
\caption{Evaluated platforms}
\begin{tabular}{@{}cccccc@{}} \toprule
\textbf{Platform} & \textbf{Type} & \textbf{Cores} & \begin{tabular}[x]{@{}c@{}}\textbf{Tech.}\\\textbf{(nm)}\end{tabular} & \begin{tabular}[x]{@{}c@{}}\textbf{Freq.}\\\textbf{(MHz)}\end{tabular} & \begin{tabular}{@{}c@{}}\textbf{Area}\\\textbf{(mm$^2$)}\end{tabular}  \\ \midrule
\begin{tabular}{@{}c@{}}Xeon\\E5-2620\end{tabular} & CPU & 6 & 32 & 2000 & 435 \\ 
%\begin{tabular}{@{}c@{}}Cortex\\A15\end{tabular} & CPU & 4 & 28 & 2300 & 90 \\ 
%Tegra Jetson K1 & GPU & 192 & 28 & 852 \\
\begin{tabular}{@{}c@{}}Titan X\end{tabular} & GPU & 3072 & 28 & 1075 & 601 \\
Kintex-7 & FPGA & N/A & 28 & 91 & 164 \\
%Xeon Phi Coproc. &  & 60 & 22 & 1053 & \\
%\begin{tabular}{@{}c@{}}Automata\\Proc.\end{tabular} & AP & 64 & 50 & \begin{tabular}{@{}c@{}}44.3-\\133\end{tabular} & \\
\tool{} & ASIC & \begin{tabular}{@{}c@{}}80-\\320\end{tabular} & 65 & 250 & \begin{tabular}{@{}c@{}}131-\\329\end{tabular}\\ \bottomrule
\end{tabular}
\label{evaluated-platforms}
\end{table}
}

\noindent\textbf{\tool{} ASIC}: To evaluate \tool{}, we implemented, synthesized, and place-and-routed our design in Verilog with the Synopsys Design Compiler and IC Compiler using a TSMC 65 nm standard cell library; SRAM memories were generated using the ARM Memory Compiler.
We also built an assembler and simulator to generate program binaries, benchmark assembly programs, and validate the correctness of our design.
To measure throughput, we use post-placement and route frequency estimates and simulate the time it takes to process each of the workloads in Table~\ref{knn-workloads}.
Each benchmark is handwritten using our instruction set defined in Table~\ref{ncam_isa}.
For power and energy efficiency estimates, we generate traces from real datasets to measure realistic activity factors.  
We then use the PrimeTime power analyzer to estimate power and multiply by the simulated run time to obtain energy efficiency estimates.
Finally, we report the area estimates provided in the post-placement and route reports normalized to a 28 nm technology.

\noindent\textbf{Xeon E5-2620 CPU}: We evaluate a six core Xeon E5-2620 as our CPU baseline. %two multicore CPUs: a Xeon E2620 and ARM Cortex A15.
For each platform, we benchmark wall-clock time using the implementations of kNN provided by the FLANN library~\cite{flann} for linear, kd-tree, and k-means based search, and the FALCONN library for hyperplane MPLSH~\cite{falconn}.
For power and energy efficiency measurements, we use an external power meter to measure dynamic compute power.
Dynamic compute power is computed by taking the difference between the load and idle power when running each benchmark.
Energy efficiency is then calculated as the product of the run time and dynamic power.
Estimates of the CPU die size is taken from~\cite{xeon-area}.

\noindent\textbf{Titan X GPU}: For our GPU comparison, we use an NVIDIA Titan X GPU using a well-optimized, off-the-shelf implementation provided by Garcia et al.~\cite{gpu_knn}.
We again record wall-clock time, and measure idle and load power using a power meter to measure run time and energy efficiency.
We estimate the die size of the Titan X from~\cite{titan-x-area}.

\noindent\textbf{Kintex-7 FPGA}: We measure the performance and energy efficiency of our implementation on a Xilinx Kintex-7 FPGA using Vivado 2014.5.
We use post-placement and route frequency estimates and simulated run times to estimate the throughput of kNN on the FPGA fabric.
For power measurements, we use the Vivado Power Analyzer tool and refer to~\cite{kintex-7-area} for device area estimates.

%%%%%%%%%%%%%%%%%%%%%%%%%%%%
% Bit Bucket
%%%%%%%%%%%%%%%%%%%%%%%%%%%%

\begin{comment}

\end{comment}

\section{Evaluation Results}
\label{sec:evaluation}

We now present throughput, power, and energy efficiency measurements of \tool{}s relative to competing heterogeneous computing platforms.
For brevity, we first evaluate Euclidean distance kNN then separately evaluate different indexing techniques and distance metrics.

\subsection{Accelerator power and area}

Our post-placement and route power and area results are shown in Figures~\ref{ncam-power-submodule} and~\ref{ncam-area-submodule} respectively for different processing unit vector lengths and different submodules in the design.
Area and power measurements are normalized to 28 nm technology using linear scaling factors.
In terms of area, a large portion of the accelerator design is devoted to the SRAMs composing the scratchpad memory.
However, relative to the CPU or GPU, the \tool{} acceleration logic is still significantly smaller.
Compared to the Xeon E5-2620, an \tool{} is 6.23-15.62$\times$ smaller while compared to the Titan X an \tool{} is 9.84-24.66$\times$ smaller.
For comparison, the die size for HMC 1.0 in ~\cite{HMC} in a 90 nm process is 729 mm$^2$; normalized to a 28 nm process, the die size would be $\approx$ 70.6 mm$^2$ which is roughly the same or larger than our \tool{} accelerator design\footnote{Die size for HMC 2.1 are not publicly available.}.
In terms of power, a \tool{} uses no more than a typical memory module which makes it compatible with the power consumption of die stacked memories.
Prior work by Puttaswamy et al.~\cite{puttaswamy06} shows temperature increases from integrating logic on die-stacked memory are not fatal to the design even for a general purpose core. 
Since \tool{} consumes less power than general purpose cores, we do not expect thermal issues to be fatal. 

{
\begin{figure*}
\begin{subfigure}{0.5\textwidth}
\centering
\includegraphics[width=\linewidth]{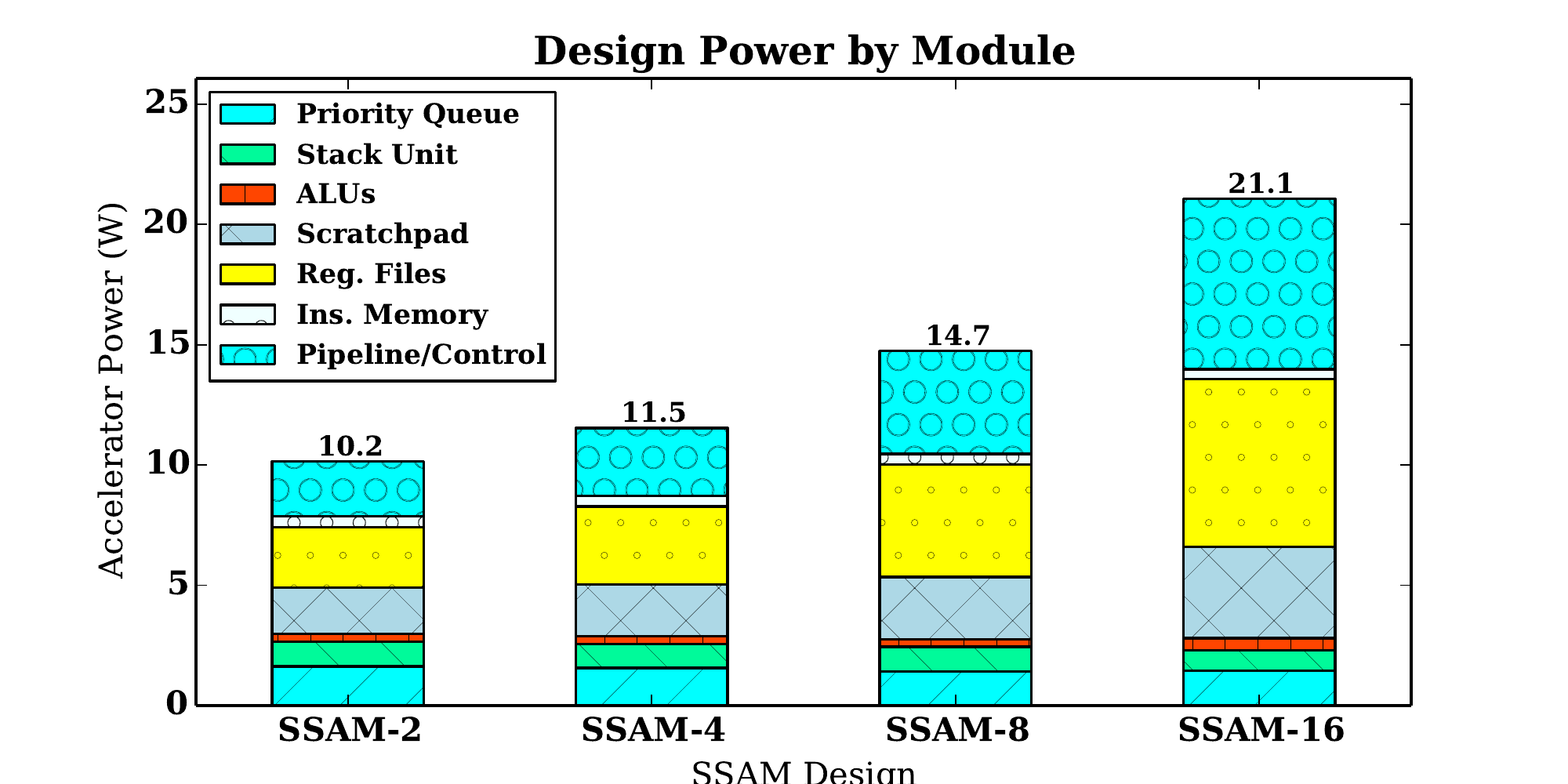}
\caption{\tool{} accelerator logic power by submodule (28 nm).}
\label{ncam-power-submodule}
\end{subfigure}%
\begin{subfigure}{0.5\textwidth}
\centering
\includegraphics[width=\linewidth]{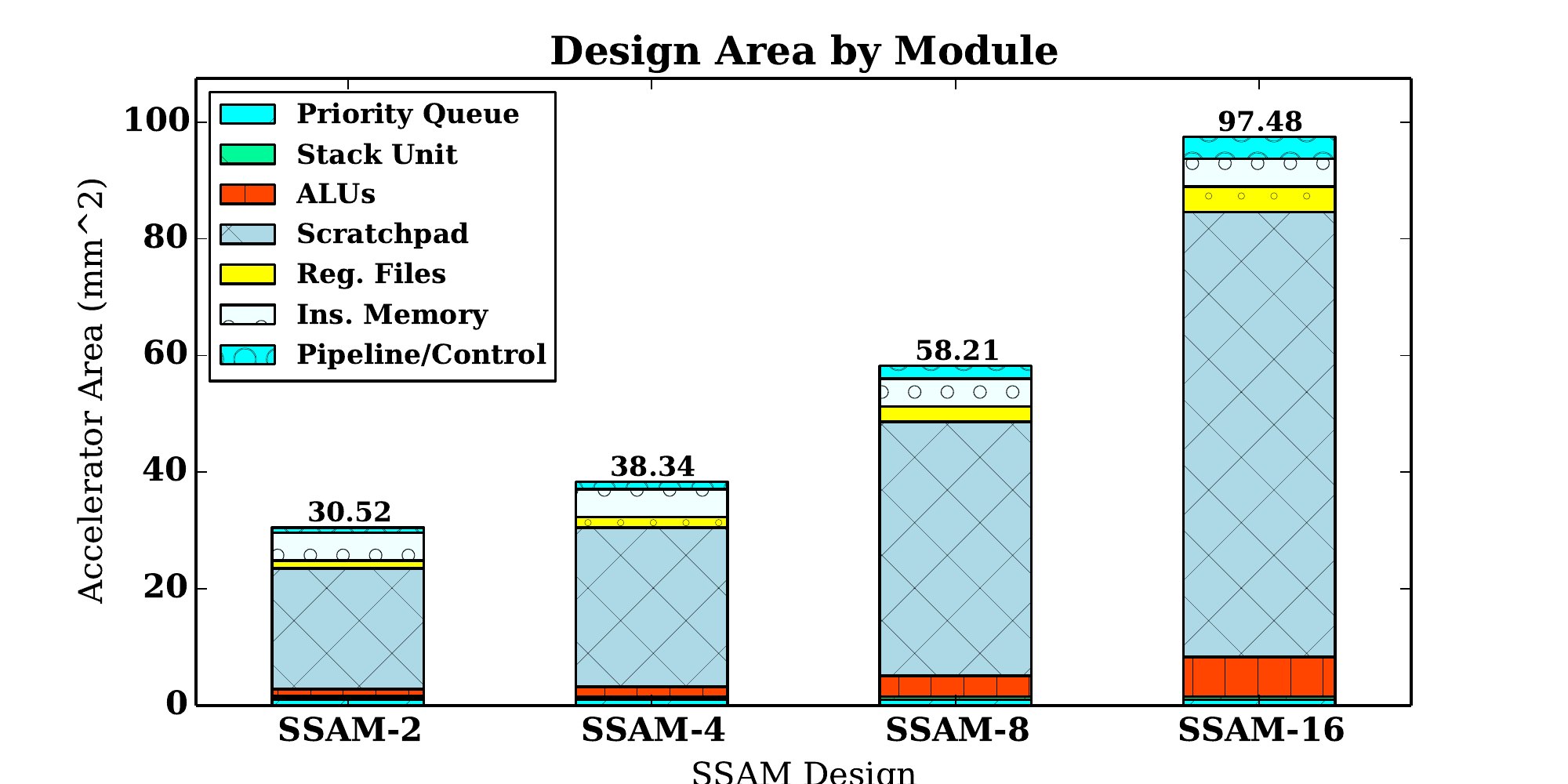}
\caption{\tool{} accelerator logic area by submodule (28 nm).}
\label{ncam-area-submodule}
\end{subfigure}
\caption{\tool{} accelerator power and area by submodule (28 nm)}
\label{power-area}
\end{figure*}
}

\subsection{Throughput and energy efficiency}

We now report area-normalized throughput and energy efficiency gains across each platform for \textit{exact linear search} which is agnostic to dataset composition and index traversal overheads. 
This quantifies the gains attributed to different heterogeneous computing technologies.
Figures~\ref{linear-results:performance} and~\ref{linear-results:energy} shows the area-normalized throughput and energy efficiency of a \tool{} against competing heterogeneous solutions.
The FPGA and \tool{} designs are suffixed by the design vector length; for instance, \tool{}-4 refers to a \tool{} design with processing units that have vector length 4.
We observe \tool{} achieve area-normalized throughput improvements of up to 426$\times$, and energy efficiency gains of up to 934$\times$ over multi-threaded Xeon E5-2620 CPU results. 
We also observe that GPUs and the FPGA implementation of the \tool{} acceleration logic exhibit comparable throughput and energy efficiency.
The FPGA in some cases underperforms the GPU since it effectively implements a soft vector core instead of a fixed-function unit; we expect that a fixed-function FPGA core would fare better.

{
\begin{figure*}[!h]
\begin{subfigure}{0.5\textwidth}
\centering
\includegraphics[width=\linewidth]{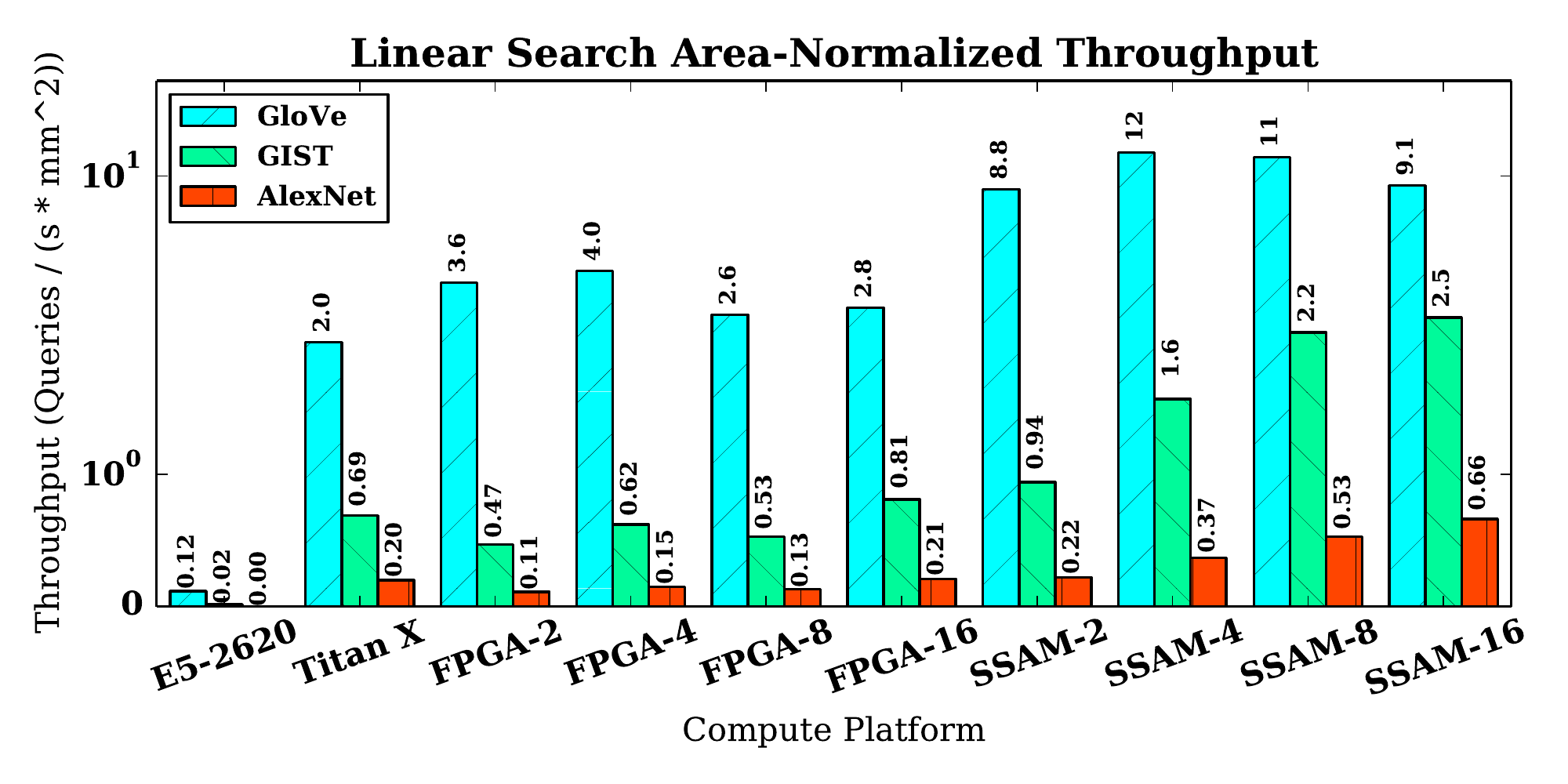}
\caption{Area-normalized Throughput}
\label{linear-results:performance}
\end{subfigure}%
\centering
\begin{subfigure}{0.5\textwidth}
\centering
\includegraphics[width=\linewidth]{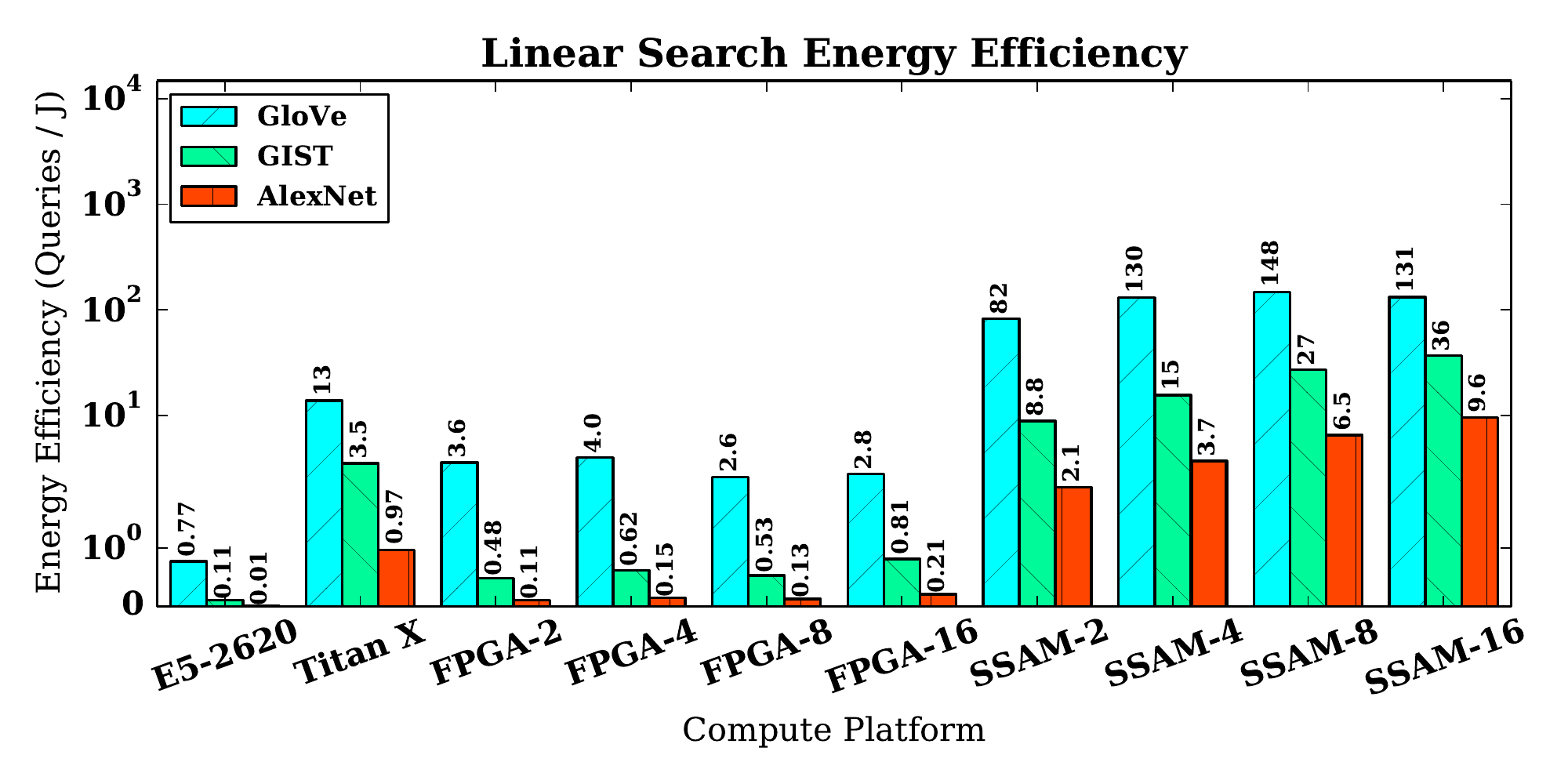}
\caption{Energy efficiency}
\label{linear-results:energy}
\end{subfigure}
\setlength{\belowcaptionskip}{-10pt}
\caption{Area-normalized throughput and energy efficiency for exact linear search using Euclidean distance.}
\label{linear-results}
\end{figure*}
}

In terms of the enhanced bandwidth, we attribute roughly one order of magnitude run time improvement to the higher internal bandwidth of HMC 2.0.
Optimistically, standard DRAM modules provide up to 25 GB/s of memory bandwidth whereas HMC 2.0 provides 320 GB/s.
For similarity search, the difference in available bandwidth directly translates to raw performance.
The remaining gains in energy efficiency and performance can be attributed mostly to accelerator specialization.
To quantify the impact of the priority queue, we simulate the performance of SSAM using a software priority queue instead of leveraging the hardware queue.
At a high level, the hardware queue improves performance by up to 9.2\% for wider vector processing units.

\subsection{Approximate kNN search}

We now evaluate the impact of approximate indexing structures and specialization on throughput and energy efficiency.
Figure~\ref{result:ncam-accuracy} compares the throughput versus accuracy curves for a \tool{} and Xeon E5-2620 CPU for each dataset.
In general, at a 50\% accuracy target we observe up to two orders of magnitude throughput improvement for kd-tree, k-means, and HP-MPLSH over CPU baselines.
The kd-tree and k-means indexing structures are still dominated by distance calculations and benefit greatly from augmented bandwidth when sequentially scanning through buckets for neighbors.
HP-MPLSH on the other hand is composed of a combination of many hash function calculations and bucket traversals; we find that for the parameter sets used in our characterization, the performance of HP-MPLSH is dominated mostly by hashing rate.
However, the parameters for HP-MPLSH can be adjusted to reduce the dependence on hash performance by reducing the number of hash bits; this would increase the number of vectors hashed to the same bucket and shift the performance bottleneck from hashing performance back to linear bucket scans.

{
\begin{figure*}[!ht]
\centering
\includegraphics[width=\linewidth]{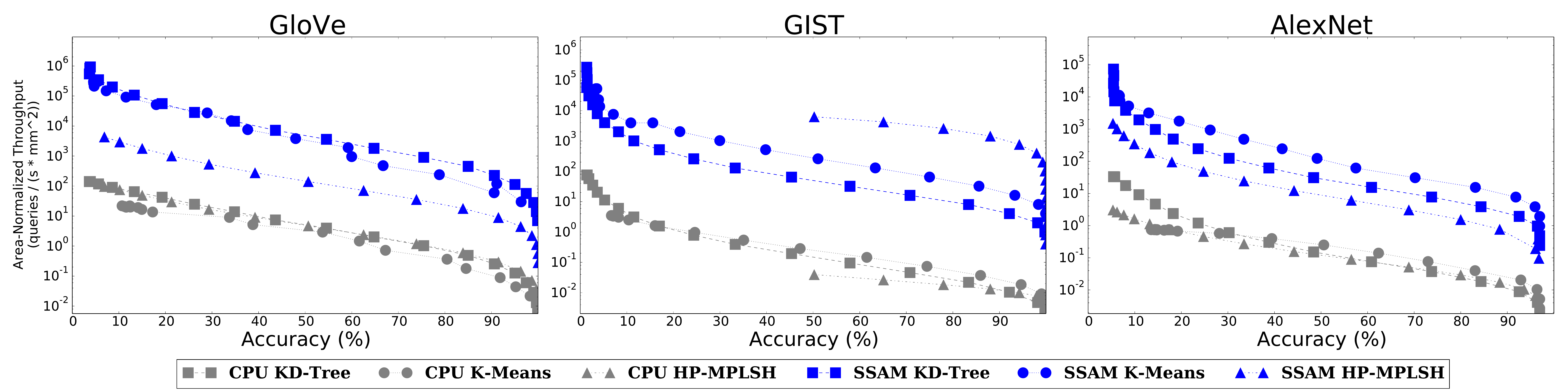}
\caption{Area-normalized throughput versus accuracy for Euclidean distance kNN using indexing structures (up and to the right is better).}
\label{result:ncam-accuracy}
\end{figure*}
}

\subsection{Alternative distance metrics}

We now briefly quantify the performance of alternative distance metrics on \tool{} for three additional distance metrics: Hamming distance, Cosine similarity, and Manhattan distance.
Unsurprisingly, the impact of binarizing data vectors and using Hamming distance provides good throughput improvement (up to 9.38$\times$) since less data must be loaded to process a vector and Hamming distances using the \code{FXP} instruction on \tool{}s are cheap.
Manhattan distance and Euclidean distances are the same cost since they require roughly the same number of operations.
Meanwhile, cosine similarity\footnote{Cosine similarity is defined as $(\sum_i{a_ib_i})^2/(\sum_i{a_i^2}\sum_i{b_i^2})$.} is about twice as expensive as Euclidean distance because of the additional numerator and divisor terms.
Fixed-point division for cosine similarity is performed in software using shifts and subtracts, however the software division is still much cheaper than the rest of the distance calculation.

\begin{table}
\centering
\caption{Relative throughput of alternate distance metrics}
\begin{tabular}{@{}cccc@{}} \toprule
Distance Metric & GloVe & GIST & AlexNet \\ \midrule
Euclidean  & 1$\times$    & 1$\times$    & 1$\times$ \\
Hamming    & 4.38$\times$ & 7.98$\times$ & 9.38$\times$ \\
Cosine similarity     & 0.46$\times$& 0.47$\times$ & 0.47$\times$ \\
Manhattan  & 0.94$\times$ & 0.99$\times$ & 0.99$\times$ \\ \bottomrule
%Jaccard    & & & \\ \bottomrule
\end{tabular}
\end{table}

%%%%%%%%%%%%%%%%%%%%%%%%%%%%%%%5
% Deprecated Text
%%%%%%%%%%%%%%%%%%%%%%%%%%%%%%%5

\begin{comment}

\end{comment}

\section{Discussion} \label{sec:discussion}

We now briefly evaluate the generality of \tool{}s for other workloads, and with respect to content addressable memories, then compare \tool{}s to alternative near-data processing technologies.

\subsection{Index construction and other applications}

The \tool{} is not limited to approximate kNN search and can also perform other data intensive operations such as index construction or data intensive applications.
In kNN, the overhead of building indexing structures is amortized away by the number of queries executed; however, index construction is still three orders of magnitude slower than single query execution.
\tool{}s can be reprogrammed to also perform these data intensive tasks; index construction also benefits from near-data processing since techniques like k-means and kd-tree construction require multiple scans over the entire dataset.
For instance, to build a hierarchical k-means index we execute k-means by treating cluster centroids as the dataset, and streaming the dataset in as kNN queries to determine the closest centroid.
While a host processor must still handle the short serialized phases of k-means, \tool{}s are able to accelerate the data intensive scans in the k-means kernel by performing the computation near memory.
Similarly for kd-tree index construction \tool{}s can be used to quickly scan the dataset and compute the variance across all dimensions; the host processor can then assign bifurcation points and generate the tree.
In both cases, the host processor must provide some control and high level orchestration but the bulk of each index construction kernel can be offloaded to exploit the benefits of high memory bandwidth.

\tool{}s can also be used to accelerate other data intensive applications that benefit from vectorization and enhanced memory bandwidth.
Applications such as support vector machines, k-means, neural networks, and frequent itemset mining can all be implemented on \tool{}.
In particular, the vectorized FXP instruction is useful for evaluation classes of application which rely on many Hamming distance calculations such as binary neural networks~\cite{binary-neural-network}, and binary hash functions.

\subsection{\tool{} as a high density generalized content addressable memory}

Semantically \tool{}s are a generalization of a content addressable memory (CAM) and ternary content addressable memory (TCAM); more importantly \tool{}s are a semantically more powerful associative computing substrate.
To use an \tool{} as a CAM, we simply fix $k=1$ and check if a neighbor has a distance of zero to the query.
To use an \tool{} as a TCAM, we use a modified Hamming distance which adjusts for ternary matches.
To do this, a ternary bit mask $T$ is supplied to an \tool{} with the query vector $Q$.
The ternary bit mask holds a 0 at positions in the query vector where a ternary match should occur and 1 otherwise.
The query vector and input data vector $D$ are XOR'ed and AND'ed with the ternary bit mask; this effectively masks off positions where the bits did not match.
We then check if the distance between $D$ and $Q$ is zero to determine if there was a ternary match.
The resulting modified Hamming distance can be expressed POPCOUNT((D$\oplus$Q) \& T).

Unlike CAMs and TCAMs, \tool{}s more generally support approximate and arbitrary width matches.
Approximate matches are more powerful since they are able to return similar content as opposed to only exact content matches.
\tool{}s can also return multiple matches while CAMs and TCAMs typically are design to only return one memory location.

Architecturally, \tool{}s realize a different design point from traditional CAMs or TCAMs since the compute units used to perform the matching are separate from the underlying memory cell implementation.
Binary CAMs and TCAMs are organized to simultaneously search all the data in parallel while \tool{}s create the illusion of an associative memory but rely on high internal bandwidth to quickly scan data and process a query.
\tool{}s also benefit from much higher bit density, capacity and reduced operating power of DRAMs while maintaining flexible associative computing capabilities.
Table~\ref{tcam-dram-comparison} provides a comparison of effective bit density between TCAM, DRAM, and \tool{} for a 16 GB capacity memory.
In terms of area, \textit{\tool{}s effectively achieve the same bit density as DRAM} because the additional acceleration logic is small relative to DRAM macro sizes.
In contrast, TCAMs are more than 19$\times$ less dense than standard DRAMs and the \tool{} design presented in this paper.
Finally, in terms of power a 20 Mb TCAM macro consumes 10.6 W~\cite{28nm-tcam} which is already significantly more than the \tool{} acceleration logic and most DRAM modules.

{
\setlength{\textfloatsep}{0.0cm}
\begin{table}[!ht]
\centering
\caption{TCAM, DRAM, \tool{} density comparison}
\label{tcam-dram-comparison}
\begin{tabular}{@{}cccc@{}} \toprule
  & \begin{tabular}{@{}c@{}}TCAM\\28nm~\cite{28nm-tcam}\end{tabular} & \begin{tabular}{@{}c@{}}DRAM\\65nm~\cite{65nm-dram}\end{tabular} & \begin{tabular}{@{}c@{}}\tool{}-4\\+DRAM\end{tabular} \\ \midrule
Density (Mb/mm$^2$)  & 0.61              & 2.17              & N/A \\
\begin{tabular}{@{}c@{}}Area @ 16GB\\(mm$^2$)\end{tabular} & 214872           & 60402                     & 60609\\
\begin{tabular}{@{}c@{}}Area @ 28nm+\\16GB (mm$^2$)\end{tabular} & 214872       & 11208                     & 11247\\
\begin{tabular}{@{}c@{}}Density @ 28nm+\\16GB(Mb/mm$^2$)\end{tabular} & \textbf{0.61}        & 11.69                     & \textbf{11.65}\\ \bottomrule
%Power         & 10.4 W / 20 Mb                                 & 0.077 W / 1 Mb                      & N/A \\
\end{tabular}
\end{table}
}

\subsection{Alternative near-data processing substrates}

Near-data processing manifests in many different shapes and forms; in this section, we briefly contrast our approach against alternative near-data processing architectures.

\noindent\textbf{Micron Automata Processor (AP)}: The AP is a near-data processing architecture specialized for high speed non-deterministic finite automata (NFA) evaluation~\cite{ap}.
Unlike \tool{}, the AP cannot efficiently implement arithmetic operations and is limited to distance metrics like Hamming distance or Jaccard similarity.
At a high level, the automata design is composed of multiple parallel NFAs where each NFA encodes a distance calculation with a single dataset vector\footnote{Details of the automata design are beyond the scope of this paper}.
A query vector is streamed into the AP and compared against all NFAs in parallel and sorted.
To support execution of different NFAs, the AP can be reconfigured much like reconfiguration on a FPGA.
We briefly evaluate the AP by designing, and compiling a design for each dataset, and use the results to build an analytical model to estimate performance for a current generation AP device.
Table~\ref{ap-comparison} shows the AP's performance and energy efficiency compared to \tool{}.
At a high level, we find that the large datasets presented in this paper do not fit on one AP board configuration, and as a result the AP is bottlenecked by the high reconfiguration overheads compared to \tool{}.

{
\begin{table}[!bt]
\centering
\caption{\tool{} and AP throughput comparison for linear Hamming distance kNN}
\label{ap-comparison}
\begin{tabular}{@{} cccc @{}} \\ \toprule
Dataset        & GloVe & GIST & AlexNet \\ \midrule
\tool{}-4 (queries/s) & 2059 & 481 & 134 \\
AP (queries/s)   & 288 & 2.64 & 0.553  \\ \bottomrule
\end{tabular}
\end{table}
}

\noindent\textbf{Compute in Resistive RAM (ReRAM)}: Computation in ReRAM is an emerging near-data processing technique that can perform limited compute operations by directly exploiting the resistive properties of ReRAM memory cells~\cite{reram-computation}.
This allows augmented computational capabilities beyond what are available to DRAM based near-data processing techniques such as \tool{}.
Most notably, Chi et al.~\cite{reram-nn} has shown how in-situ ReRAM computation can be used to accelerate convolutional neural networks without moving data out of the ReRAM cells themselves.
As the technology matures, it would not be unprecendented to replace DRAM in favor of ReRAM and its augmented computing capabilities.

\noindent\textbf{In-Storage Processing}: There has also been a renewed interest in instantiation computation near disk or SSD.
Recent work such as Intelligent SSD~\cite{intelligent-ssd, case-for-intelligent-ssd} and Biscuit~\cite{biscuit} have all proposed adding computation near mass storage devices and shown promising improvements for applications like databases.
However, compared to \tool{}, in-storage processing architectures target a different bandwidth to storage capacity design point.
Unlike \tool{}, SSD-based near-data processing handles terabytes of data at lower bandwidth speeds which is less ideal for latency critical applications like similarity search.

\noindent\textbf{Die-Stacked HMC Architectures (This Paper)} Instantiating an accelerator adjacent to HMC is not a new proposal~\cite{ahn15, amc, in-memory-nn}; prior work has shown that such an architectural abstraction is useful for accelerating graph processing~\cite{ahn15} and neural networks~\cite{in-memory-nn}. %\bluecomment{TODO: buff citations}
This architecture has several advantages over in-situ ReRAM computation and the automata processor.
First, by abstracting the computation away from the memory substrate, the types of computation supported is decoupled from the restrictions of underlying memory implementations.
Second, by separating the computation from actual memory cells, this architectural abstraction achieves much higher compute and memory density; this is unlike substrates like the AP where compute and memory are both instantiated in the same memory process.

%%%%%%%%%%%%%%%%%%%%%%%%%%%%%%%
% Bit Bucket
%%%%%%%%%%%%%%%%%%%%%%%%%%%%%%%

\begin{comment}

\end{comment}

\section{Related Work}\label{sec:related-works}
The concept of near-data processing has been studied in the literature for decades. 
More interestingly, the concept of integrating similarity search accelerators with memory also has an established history, indicating ongoing interest.

\noindent\textbf{CAMs:} As far back as the 1980s and 1990s, near-memory accelerators were proposed to improve the performance of nearest neighbor search using CAMs~\cite{roberts90}.
Kohonen et al.~\cite{kohonen89} proposed using a combination of CAMs and hashing techniques to perform nearest neighbor search. 
Around the same time, Kanerva et al.~\cite{kanerva88} propose sparse distributed memory (SDM) and a ``Best Match Machine'' to implement nearest neighbor search.
The ideas behind SDM were later employed by Roberts in PCAM~\cite{roberts90} which is, to the best of our knowledge, the first fabricated CAM-based accelerator capable of performing nearest neighbor search on its own. 

Algorithms that exploit TCAMs to perform content addressable search such as ternary locality sensitive hashing (TLSH)~\cite{shinde10} and binary-reflected Gray code~\cite{bremler-barr15} do exist. 
However, TCAMs suffer from lower memory density, higher power consumption, and smaller capacity than emerging memory technologies.  
While prior work~\cite{guo11} shows promising increases in performance, energy efficiency, and capacity, TCAM cells are less dense than DRAM cells. 
For the massive scale datasets in kNN workloads, the density disparity translates to an order of magnitude in cost. 
Despite these limitations, there is still active work in TCAMs for data-intensive applications to accelerator associative computations~\cite{guo13}.

\noindent\textbf{Multiprocessor and Vector PIMs:}
In the late 1990s, Patterson et al.~\cite{patterson97} proposed IRAM which introduced processing units integrated with DRAM.
In particular, Gebis et al.~\cite{viram1am} and Kozyarakis et al.~\cite{kozyarakis-viram} proposed VIRAM which used a vector processor architecture with embedded DRAM similar to our work. 
Similar to our work, the intention of VIRAM was to capitalize on the higher bandwidth and reduce energy consumption by co-locating general MIPS cores and vector register and compute units near DRAM.
Unlike VIRAM, \tool{} does not implement a full cache hierarchy, targets a different class of algorithms, and uses a 3D die-stacked solutions.

Kogge et al.~\cite{kogge95} propose the EXECUBE architecture which integrates general purpose cores with DRAM macros. 
Elliott et al.~\cite{elliott99} propose C-RAM which add SIMD processing units adjacent to the sense amplifiers capable of bit serial operations. 
Active Pages~\cite{activepages} and FlexRAM~\cite{flexram} envisioned a programmable processing element near each DRAM macro block which could be programmed for kNN acceleration. 
However, none of these prior efforts directly addresses the kNN search problems we discuss.

More recently, Active Memory Cube (AMC)~\cite{amc} proposes a similar vector processing unit and cache-less system on top of HMC.
While both \tool{} and AMC arrive at the same architecture conclusion - that vector PIM on die-stacked DRAM is useful - our work provides a more application-centric approach which allows us to codesign architectural features such as the priority queue.

\noindent\textbf{Application-Driven PIM:} Application-justified PIM design is not a new idea.
Deering et al.~\cite{deering94} propose FBRAM, a ``new form'' of memory optimized to improve random access writes to accelerate z-buffering for 3D graphics algorithms.  
Lipman and Yang~\cite{Lipman97thesmart} propose a DRAM based architecture called smart access memory (SAM) for nearest-neighbor search targeting DB applications. 
Their design tightly integrates a k-nearest neighbor accelerator engine and microarchitecturally shares common elements with our design. 
Agrawal et al.~\cite{lebeck_nn} exploit accelerators to reduce the total cost of ownership of high-dimensional similarity search. 
Yu et al.~\cite{Yu_nn} optimize all-pairs nearest neighbors by fusing neighbor selection with distance computations. 
Finally, Tandon et al~\cite{tandon13} propose an all pairs similarity accelerator for NLP; however, their work integrates their accelerator with the last level cache instead of memory.

The emergence and maturity of die-stacked memory and alternative memory substrates has also enabled a wide variety of PIM accelerator proposals~\cite{chi16, ahn15, amc, hsieh16, hsieh16iccd, pim-taxonomy, pim-graphics, top-pim, in-memory-nn}.
Chi et al.~\cite{chi16}, Kim et al.~\cite{kim16}, and Gao et al.~\cite{gao17} all propose PIM solutions for accelerating neural networks.
Ahn et al.~\cite{ahn15} propose PIM on top of HMC for graph processing, and Hsieh et al.~\cite{hsieh16} and Zhang et al.~\cite{zhang14} propose PIM-based GPU architectures.
Imani et al.~\cite{imani17} propose MPIM for linear kNN search; however, their architecture uses a resistive RAM-based approach and is limited to bitwise operations. 
Furthermore, MPIM does not consider modern approximate kNN indexing algorithms nor does it evaluate the quality versus accuracy tradeoffs that these algorithms make.

\section{Conclusions}
\label{sec:conclusions}

We presented \tool{}, an application-driven near-data processing architecture for similarity search.
We showed that by moving computation closer to memory, \tool{} is able to address the data movement challenges of similarity search and exploit application codesign opportunities to accelerate similarity search.
While we used HMC as our memory backend, the high-level accelerator design and insights still generalize to alternative memory technology and in-memory processing architectures.
The PIM proposal presented in this paper are also relevant to other data intensive workloads where data movement is becoming an increasingly fundamental challenge in improving system efficiency. 
%Our work is one instantiation of many opportunities where PIM can provide orders of magnitude throughput and energy efficiency improvements.

%%%%%%% -- PAPER CONTENT ENDS -- %%%%%%%%

%%%%%%%%% -- BIB STYLE AND FILE -- %%%%%%%%
\balance
\bibliographystyle{ieeetr}
\bibliography{references}
%%%%%%%%%%%%%%%%%%%%%%%%%%%%%%%%%%%%

\end{document}